\documentstyle[preprint,aps,tighten]{revtex}
\begin{document}
\draft

\def\pp{{\, \mid \hskip -1.5mm =}}
\def\cL{{\cal L}}
\def\be{\begin{equation}}
\def\ee{\end{equation}}
\def\bea{\begin{eqnarray}}
\def\eea{\end{eqnarray}}
\def\beaa{\begin{eqnarray*}}
\def\eeaa{\end{eqnarray*}}
\def\tr{{\rm tr}\, }
\def\nn{\nonumber \\}
\def\e{{\rm e}}
\def\D{{D \hskip -3mm /\,}}

\title{{\small\hfill OKHEP-00-11}\\
Quantum (In)Stability of a Brane-World AdS$\bf_5$ 
Universe at Nonzero Temperature}

\author{Iver Brevik\thanks{Electronic address: iver.h.brevik@mtf.ntnu.no}}
\address{Division of Applied Mechanics, 
Norwegian University of Science and Technology,
N-7491 Trondheim, NORWAY}

\author{Kimball A. Milton\thanks{Electronic address: milton@mail.nhn.ou.edu}}
\address{Department of Physics and Astronomy, The University of Oklahoma,
Norman, OK 73019 USA}

\author{Shin'ichi Nojiri\thanks{Electronic address: nojiri@cc.nda.ac.jp}}
\address{Department of Applied Physics,
National Defence Academy,
Hashirimizu Yokosuka 239-8686, JAPAN}

\author{Sergei D. Odintsov\thanks{On leave from Tomsk State Pedagogical
 University, 
634041 Tomsk, RUSSIA.
Electronic address: odintsov@ifug5.ugto.mx}}

\address{
Instituto de Fisica de la Universidad de Guanajuato,
Lomas del Bosque 103, Apdo.\ Postal E-143, 
37150 Leon, Gto., MEXICO }
\date{\today}
\maketitle
\begin{abstract}  

We consider the quantum effects of bulk matter (scalars, spinors)
in the Randall-Sundrum AdS$_5$ brane-world at nonzero temperature. 
The thermodynamic energy (modulus potential) is evaluated at low and 
high temperatures. This potential has an extremum which could be a minimum 
in some cases (for example, for a single fermion). That suggests a
 new dynamical mechanism to stabilize the 
thermal AdS$_5$ brane-world. It is shown that the brane separation required  
to solve the hierarchy scale problem may occur at a quite low temperature.
A natural generalization in terms of the AdS/CFT correspondence (through 
the supergravity thermal contribution) is also possible.

\end{abstract}


\section{Introduction}

It is expected that the Randall-Sundrum scenario \cite{RS} (or warped 
compactification) should be realized completely in terms of the 
anti-deSitter/conformal field theory (AdS/CFT)
correspondence \cite{AdS}. There are various proposals for 
such a realization. For example, taking  the quantum 
effects of brane CFT (via 
the corresponding conformal anomaly) may help in the construction of
new brane-worlds \cite{NOZ,HHR,ANO,NOO} which occur in the AdS/CFT 
correspondence (as a kind of holographic renormalization group (RG) 
flow). It is remarkable 
that a calculation of corrections to Newton's law from quantum brane CFT
coincides quite well with that computed from supergravity \cite{DL}.

The role played by quantum effects suggests that they may be 
relevant in other aspects of warped (brane-world) 
compactifications. Indeed, for usual two-dimensional or four-dimensional AdS 
backgrounds the matter quantum effects may help to
stabilize (or destabilize) the AdS background under 
consideration \cite{NOZ,BO}. In the same way, one can expect that 
quantum bulk gravity may be important in the realization of warped 
compactification within the AdS/CFT scheme or in improving the
Randall-Sundrum scenario.
As a toy model for quantum bulk gravity one can consider 
a quantum bulk scalar \cite{GPT,NOZ1,GR,KKS}. 

As has been suggested in Refs.~\cite{GPT,GR} bulk quantum 
effects may generate a modulus effective potential which 
will help in stabilizing the brane-world five-dimensional AdS space with
a finite separation of (flat) branes,
which also solves the hierarchy problem. Unfortunately, the explicit 
analysis done in Ref.~\cite{GR} exhibits a power suppression of 
quantum bulk effects, which consequently have a negligible role in 
stabilizing the brane-world scenario with the necessary 
hierarchy. (In the original RS scenario this purpose is achieved by
fine-tuning of parameters.)

Nevertheless, it may happen that, if we take into account
other effects or a change of topology \cite{MI},
bulk quantum fields may provide the 
dynamical mechanism to construct the stable AdS brane-world 
without fine-tuning.
As one step in this direction, we consider the quantum 
effects of the bulk 
 matter in a five-dimensional AdS universe at nonzero temperature.
The corresponding high- and low-temperature limits of the 
thermodynamic energy are
estimated and applied to the study of quantum stability of  
an AdS$_5$ (flat) brane-world. It is demonstrated that the thermal modulus 
potential may have an extremum which can be a minimum in some cases.
Hence, a new thermal dynamical mechanism to stabilize a five-dimensional 
AdS brane-world is suggested. It is interesting that the brane separation 
required to solve the hierarchy problem may occur as a minimum 
of the effective potential at a quite low temperature.

\section{Effective potential in 5-dimensional
 A\lowercase{d}S space at nonzero temperature}

Our purpose in this section will be the calculation of the energy 
(effective potential) for a bulk quantum field on a five-dimensional AdS 
background at nonzero temperature.
We start with a textbook-like review of black body 
radiation. Since the energy of a single particle with momentum 
${\bf p}$ and mass $m$ is given by
\be
\label{b1}
E_p=\sqrt{{\bf p}^2 + m^2}\ ,
\ee
we obtain the following expression for its contribution to 
the partition function $Z$
\begin{mathletters}
\be
\label{b2}
Z_p^b=\sum_{n=0}^\infty\e^{-\beta E_p\left( n+{1 \over 2}\right)}
={1 \over 2\sinh \left({\beta E_p \over 2}\right)}\ ,
\ee
for a boson and 
\be
\label{b3}
Z_p^f=\sum_{n=0}^1\e^{-\beta E_p\left( n - {1 \over 2}\right)}
=2\cosh \left({\beta E_p \over 2}\right)\ ,
\ee
\end{mathletters}
for a fermion. Here $\beta$ is the inverse of the temperature $T$,
$\beta= 1/T$, and we sum over all possible states with $n$ 
particles (quanta). In Eqs.~(\ref{b2}) and (\ref{b3})  we
express the zero-point energy by $\pm{1 \over 2}E_p$. 
The total partition function of the particle with mass $m$ 
in a $d-1$-dimensional volume $V_{d-1}$ is given 
by summing over $Z_k$ with respect to the momentum ${\bf p}$:
\begin{mathletters}
\bea
\label{b4}
\beta F^b&=& -\ln Z^b = V_{d-1}\int {d^{d-1}{\bf p} \over 
(2\pi)^{d-1}}\ln \left(2\sinh \left({\beta E_p \over 2}\right)
\right)\ ,\\
\label{b5}
\beta F^f&=& -\ln Z^f = -V_{d-1}\int {d^{d-1}{\bf p} \over 
(2\pi)^{d-1}}\ln \left(2\cosh \left({\beta E_p \over 2}\right)
\right)\ .
\eea
Here $F^{b,f}$ is the free energy. 
 Eqs.~(\ref{b4}) and (\ref{b5}) diverge, and require 
regularization in general. For 
the supersymmetric case, we obtain a finite result:
\be
\label{b6}
\beta F^s = \beta\left(F^b + F^f\right)
= V_{d-1}\int {d^{d-1}{\bf p} \over 
(2\pi)^{d-1}}\ln \left(\tanh \left({\beta E_p \over 2}\right)
\right)\ .
\ee
\end{mathletters}
The average energy $E$ is given by the derivative of the free energy,
\be
\label{e1}
E={\partial \over \partial \beta}\left(\beta F\right)\ .
\ee
Then for  above cases (\ref{b4}), (\ref{b5}) and (\ref{b6}), 
we find
\begin{mathletters}
\bea
\label{e2}
E^b_d(\beta)&=& V_{d-1}\int {d^{d-1}{\bf p} \over 
(2\pi)^{d-1}} {E_p \over 2} \coth \left(
{\beta E_p \over 2}\right)\ ,\\
\label{e3}
E^f_d(\beta)&=& -V_{d-1}\int {d^{d-1}{\bf p} \over 
(2\pi)^{d-1}}{E_p \over 2} \tanh \left({\beta E_p \over 2}\right)
\ ,\\
\label{e4}
E^s_d(\beta) &=& V_{d-1}\int {d^{d-1}{\bf p} \over 
(2\pi)^{d-1}} {E_p \over \sinh \beta E_p} \ .
\eea
\end{mathletters}
This completes our elementary review of quantum statistics
at nonzero temperature.

In the Randall-Sundrum model \cite{RS}, the Standard Model fields are 
confined on one of the two 3-branes, which are the boundaries 
of the bulk AdS$_5$ space, whose metric is given by 
\be
\label{b7}
ds^2=-r_c^2 d\phi^2 + \e^{-2kr_c|\phi|} \eta_{\mu\nu}
dx^\mu dx^\nu\ .
\ee
Here the coordinate $\phi$ takes a value in $[-\pi,\pi]$,
$k$ is a parameter of order the Planck scale, and $r_c$ is the length of
the orbifold.
 For simplicity, let us consider the bulk scalar field $\Phi$, whose action is 
given by
\bea
\label{b8}
S_\Phi={1 \over 2}\int d^4x \int_{-\pi}^\pi d\phi \sqrt{-G}
 \left\{G^{AB}\partial_A \Phi \partial_B \Phi 
 - \left(m^2 + {2\alpha k \over r_c}\left(\delta(\phi)
 - \delta(\phi - \pi)\right)\right)\Phi^2 \right\}\ ,
\eea
where $\alpha$ parametrizes the mass on the boundaries.
All fields in the bulk 5-dimensional spacetime can be regarded as 
 Kaluza-Klein modes, which in their own right can be considered as 4-dimensional
fields on the brane with an infinite tower of masses. 
The mass spectrum $m_n$ of the Kaluza-Klein modes in $\Phi$ 
is given \cite{GW,GP} by roots of 
\be
\label{b9}
j_\nu(x_n)y_\nu(ax_n) - j_\nu(ax_n)y_\nu(x_n)=0\ ,
\ee
where $x_n=m_n/ak$, $a=\e^{-kr_c\pi}$, 
$\nu=\sqrt{4 + {m^2 \over k^2}}$, and the altered Bessel functions are
\be
j_\nu(z)=(2-\alpha)J_\nu (z) + zJ_\nu'(z),\quad
y_\nu(z)=(2-\alpha)Y_\nu (z) + zY_\nu'(z).
\ee
The radius $r_c$ can be regarded as the vacuum expectation 
value of the radion field $T(x)$. In the following, we will be
 interested in the effective potential (energy) for $a$.

 If we assume that  a function $f(x)$ is not singular for real 
positive $x$, while another function $g(x)$ has first-order zeroes 
when $x=x_n>0$, $(n=1,2,\cdots)$, then we obtain by an elementary
argument the following formula
\be
\label{b10}
\sum_n f(x_n)=-{1 \over 2\pi i}\int_C dz f'(z) 
\ln g(z)\ .
\ee
Here the contour $C$ can be taken to encircle the positive real axis,
\be
\label{b11}
\int_C dz \cdots 
= \left(\int_{+\infty -i\epsilon}^{\epsilon' -i\epsilon}
+ \int_{\epsilon' -i\epsilon}^{\epsilon' +i\epsilon}
+ \int_{\epsilon' +i\epsilon}^{+\infty +i\epsilon}\right)
dz\cdots \ ,
\ee
where $\epsilon$ and $\epsilon'$ are infinitesimally small 
positive constants. 

 Using Eq.~(\ref{b10}), the total free energy can 
be obtained by summing up the KK modes given by Eq.~(\ref{b9}):
\begin{mathletters}
\bea
\label{b12}
\beta F^{bKK}&=&{\cal F}\left[\ln\left( 2\sinh \left(
{\beta \sqrt{{\bf p}^2 + a^2k^2x^2} \over 2}\right)\right)\right],
\\
\label{b13}
\beta F^{fKK}&=&-{\cal F}\left[\ln \left(2\cosh 
\left({\beta \sqrt{{\bf p}^2 + a^2k^2x^2} 
\over 2}\right)\right)\right],
\\
\label{b14}
\beta F^{sKK}&=&{\cal F} \left[\ln 
\tanh \left({\beta \sqrt{{\bf p}^2 + a^2k^2x^2} 
\over 2}\right)\right] \ ,
\eea
\end{mathletters}
where the functional ${\cal F}$ is defined by
\be
{\cal F}[f(p,x)]=V_{d-1}\int {d^{d-1}{\bf p} \over (2\pi)^{d-1}}
{i\over2\pi}\int_C dx {d \over dx}f(p,x)
\ln\left(j_\nu(x)y_\nu(ax) - j_\nu(ax)y_\nu(x)\right)\ . 
\ee
The corresponding energies are given by
\begin{mathletters}
\bea
\label{e5}
E^{bKK}&=&{\cal F}\left[{\sqrt{{\bf p}^2 + a^2k^2x^2} \over 2}
\coth \left({\beta \sqrt{{\bf p}^2 + a^2k^2x^2} 
\over 2}\right)\right]\ ,\\
\label{e6}
E^{fKK}&=&-{\cal F}\left[
{\sqrt{{\bf p}^2 + a^2k^2x^2} \over 2}
\tanh \left({\beta \sqrt{{\bf p}^2 + a^2k^2x^2} 
\over 2}\right)\right] \ ,\\
\label{b14b}
E^{sKK}&=&{\cal F}\left[
{\sqrt{{\bf p}^2 + a^2k^2x^2} \over 
\sinh \left(\beta \sqrt{{\bf p}^2 + a^2k^2x^2} \right)}\right] \ .
\eea
\end{mathletters}

It is interesting that one can consider the thermodynamic
correction coming from the quantum radion 
itself by starting from the following Lagrangian \cite{GR}:
\be
\label{b15}
{\cal L}={f^2 \over 2}\left(\partial a\right)^2
 - \delta V_v a^4 \ .
\ee
Here  we regard $a$ as the dynamical field $a=\e^{-k\pi T(x)}$ 
 expressed in terms of the radion field $T(x)$, while
$\delta V_v$ is a small classical shift in the TeV brane 
tension relative to the value which generates the background 
metric. Note that $f$ is given  in 
terms of the five-dimensional Planck scale $M_5$ as
$f=\sqrt{24M_5^3/k}$.
If one divides $a$ into a sum of background and fluctuation parts by
\be
\label{b16}
a\rightarrow a + {1 \over f}\delta a\ ,
\ee
the Lagrangian (\ref{b15}) assumes the following form
\be
\label{b17}
{\cal L}={1 \over 2}\left(\partial \delta a\right)^2
 - \delta V_v \left(a^4 + {4 \over f}a^3\delta a 
 + {6 \over f^2}a^2 \delta a^2 + {4 \over f^3}a \delta a^3
 + {1 \over f^4}\delta a^4\right)\ .
\ee
Then the effective mass $\hat m$ is given by 
\be
\label{b18}
\hat m^2= {12 \delta V_v a^2\over f^2}\ 
\ee
and from Eqs.~(\ref{b1}), (\ref{b4}), and (\ref{e2}), the free energy and 
energy have the following forms:
\bea
\label{b19}
\beta F^r &=& V_{d-1}\int {d^{d-1}{\bf p} \over 
(2\pi)^{d-1}}\ln \left(2\sinh \left({\beta 
\sqrt{{\bf p}^2 + {12 \delta V_v a^2 \over f^2} } \over 2}\right)
\right) \ ,\\
\label{b20}
E^r &=& V_{d-1}\int {d^{d-1}{\bf p} \over 
(2\pi)^{d-1}}{\sqrt{{\bf p}^2 + {12 \delta V_v a^2 \over f^2} } 
\over 2} \coth \left({\beta 
\sqrt{{\bf p}^2 + {12 \delta V_v a^2 \over f^2} } \over 2}\right)\ .
\eea
Similarly, the contribution of other (higher spin) bulk fields 
may be taken into account. 

\section{High Temperature Limit}

 Let us consider the high temperature limit where $\beta$ is 
small. Since the zero temperature contributions have been 
evaluated in Ref.~\cite{GR}, we subtract these contributions 
from $E^b_d(\beta)$ in Eq.~(\ref{e2}) and $E^f_d(\beta)$ 
in Eq.~(\ref{e3}). ($E^s_d(\beta)$ in Eq.~(\ref{e4}) does not  contain
a zero temperature contribution):
\be
\label{b21}
\tilde E^b_d(\beta) = E^b_d(\beta) - E^b_d(\infty) \ , \quad
\tilde E^f_d(\beta) = E^f_d(\beta) - E^f_d(\infty) \ .
\ee
We should note that $\tilde E^b_d(\beta)$ and 
$\tilde E^f_d(\beta)$ are finite. If we change the momentum variable
${\bf p}$ to
\be
\label{b21b}
|{\bf p}|={q \over \beta}\ ,
\ee
Eq.~(\ref{b21}) may be rewritten as follows
\bea
\label{b21c}
\tilde E^{b,f}_d(\beta) ={V_{d-1} \over 2^{d-2}\pi^{d-1 \over 2}
\Gamma\left({d-1 \over 2}\right)\beta^d}
\int_0^\infty dq\, q^{d-2} {\sqrt{q^2 + \beta^2 m^2} \over 
\e^{\sqrt{q^2 + \beta^2 m^2}} \mp 1} \ .
\eea
When $d=4$, the expressions in Eq.~(\ref{b21c}) can be expanded with 
respect to $\beta$ as follows:
\bea
\label{b21d}
\tilde E^{b,f}_4(\beta) = {V_3 \over 2\pi^2 \beta^4}
\int_0^\infty dq \left\{ {q^3 \over \e^q \mp 1} 
+ \beta^2 m^2 \left( {q \over 2\left(\e^q \mp 1\right)} 
- {q^2\e^q \over 2\left(\e^q \mp 1\right)^2} \right) 
+ {\cal O}\left(\beta^4\right) \right\}\ .
\eea
Using the well-known formulas
\bea
\label{b21e}
\int_0^\infty {x^{s-1} \over \e^x - 1}dx = \Gamma(s)\zeta(s)\ , \quad
\int_0^\infty {x^{s-1} \over \e^x + 1}dx &=& (1-2^{1-s})\Gamma(s)\zeta(s)\ , 
\eea
we obtain
\begin{mathletters}
\label{26}
\bea
\label{b22}
\tilde E^b_4(\beta)&=&{V_3 \over 2\pi^2 \beta^4}
\left( {\pi^4 \over 15} - {\pi^2 \beta^2 m^2 \over 12} 
+ {\cal O}\left(\beta^4\right) \right)\ , \\ 
\label{b23}
\tilde E^f_4(\beta)&=&{V_3 \over 2\pi^2 \beta^4}
\left(  {7\pi^4 \over 120} - {\pi^2 \beta^2 m^2 \over 24} 
+ {\cal O}\left(\beta^4\right) \right)\ , \\ 
\label{b24}
E^s_4(\beta)&=&\tilde E^b_d(\beta) + \tilde E^f_d(\beta) 
={V_3 \over 2\pi^2 \beta^4}
\left( {\pi^4 \over 8} - {\pi^2 \beta^2 m^2 \over 8} 
 - {\cal O}\left(\beta^4\right) \right) \ .
\eea
\end{mathletters}
The first term corresponds to Wien's law for black body 
radiation of massless fields. 

In order to sum up the contribution from the Kaluza-Klein modes, 
we use the following formula from Ref.~\cite{GR} (which is obtained from
Eq.~(\ref{b10}) by unfolding $C$ so that it lies along the imaginary axis):
(the large $t$ behavior for $a\ll1$ has been removed)
\bea
\label{b25}
A_s(a)&\equiv& \sum_n x_n^{-s} \nn
&=& {s \over \pi}\sin\left({\pi s \over 2}\right)
\int_0^\infty dt\,t^{-s-1}  \ln \left[{2 \over t\sqrt{a} }
\e^{-t(1-a)}\left\{k_\nu (t) i_\nu (at) - k_\nu (at) i_\nu (t)
\right\}\right]
\eea
for $s=0$, $-2$. (This expression may be obtained on the basis of 
the zeta-regularization technique, see Ref.~\cite{EORBZ}). 
Here the Bessel functions of imaginary argument are
\bea
\label{b25b}
i_\nu(z)&=&(2-\alpha)I_\nu(z) + zI_\nu'(z) \nn
k_\nu(z)&=&(2-\alpha)K_\nu(z) + zK_\nu'(z)\ .
\eea
($I_\nu(z)$ and 
$K_\nu(z)$ are modified Bessel functions). 
The expression in Eq.~(\ref{b25}) is valid when $-1<s<0$ and we now 
consider its analytic continuation to $s=0$ or $s=-2$. 

When $s=0$, the integration in Eq.~(\ref{b25}) contains a divergence 
coming from the integration when $t\sim 0$ because
\bea
\label{bb0}
&& \ln \left[{2 \over t\sqrt{a} }
\e^{-t(1-a)}\left\{k_\nu (t) i_\nu (at) - k_\nu (at) i_\nu (t)
\right\}\right] \nn
&& \quad \sim \ln \left[{(2-\alpha -\nu)(2-\alpha+\nu) \over\nu t \sqrt{a}}
\left(a^{-\nu}-a^{\nu}\right)\right]\ .
\eea
Thus we divide the region of 
the integration into two parts ($\epsilon\ll1$)
\be
\label{bb1}
\int_0^\infty \cdots \rightarrow \left(\int_0^\epsilon 
+ \int_\epsilon^\infty\right)\cdots\ ,
\ee
and  rewrite Eq.~(\ref{b25}) in the following form:
\bea
\label{bb2}
A_s(a)
&=& {s \over \pi}\sin\left({\pi s \over 2}\right)
\left(A^{(1)}_s(a) + A_s^{(2)}(a)\right)\ . 
\eea
When $s\rightarrow 0-$, $A_s^{(1)}(a)$ is finite but $A_s^{(2)}(a)$ 
behaves as 
\be
\label{bb3}
A_s^{(2)}(a)\sim-
 \int_0^\epsilon dt\,t^{-s-1}\ln t\sim {\epsilon^{-s} \over s^2}\ .
\ee
Then in the limit of $s\rightarrow 0-$, we obtain
\be
\label{bb4}
A_0(a)={1 \over 2}\ .
\ee

On the other hand, when $s\rightarrow -2$, the divergence of 
the integration in Eq.~(\ref{b25}) comes from $t\rightarrow \infty$ 
since
\bea
\label{bb5}
&& \ln \left[{2 \over t\sqrt{a} }
\e^{-t(1-a)}\left\{k_\nu (t) i_\nu (at) - k_\nu (at) i_\nu (t)
\right\}\right]
\sim \left({13\over8}-\alpha-{\nu^2\over2}
\right)\left(1 - {1 \over a}\right)t^{-1} \nn
&& \quad \ + \left({\nu^2\over4}-{\alpha^2\over2}+{3\alpha\over2}-{19\over16}
\right)\left(1 + {1 \over a^2}\right) t^{-2}\equiv h(t)\ .
\eea
Then we divide the region of 
the integration into two regions ($\Lambda\gg1$)
\be
\label{bb6}
\int_0^\infty \cdots \rightarrow \left(\int_0^\Lambda 
+ \int_\Lambda^\infty\right)\cdots\ ,
\ee
and  rewrite Eq.~(\ref{b25}) in the following form:
\bea
\label{bb7}
A_s(a)
&=& {s \over \pi}\sin\left({\pi s \over 2}\right)
\left(\hat A^{(1)}_s(a) + \hat A_s^{(2)}(a)\right)\ , \nn
\hat A_s^{(2)}(a)&\equiv& \int_\Lambda^\infty dt\,t^{-s-1} h(t). 
\eea
One should note that $\hat A_s^{(1)}(a)$ is finite when $-3<s<0$. 
When $-1<s<0$, the integrals appearing in 
$\hat A_s^{(2)}(a)$ have the following form:
\be
\int_\Lambda^\infty dt\,t^{-s-2}={\Lambda^{-s-1}\over s+1},\quad
\int_\Lambda^\infty dt\,t^{-s-3}={\Lambda^{-s-2}\over s+2}.
\ee
Then by analytically continuing to $s\rightarrow -2$, one obtains
\bea
\label{bb9}
A_{-2}(a)=
\left[{\nu^2\over4}-{\alpha^2\over2}+{3\alpha\over2}-{19\over16}\right]
\left(1 + {1 \over a^2}\right) \ .
\eea

An important remark is in order. The same technique may be applied 
to the calculation of the contribution to the energy from higher 
spin fields (vectors, tensors) \cite{GPT} at nonzero temperature.
Only the corresponding values for $\alpha$, $\nu$ and the  mass $m$ will be 
changed.

The energy $E(a)$ can be regarded as an effective potential 
with respect to the expectation value of the radion field or $a$, 
which determines the distance between the two branes.  
Note that there appears a nontrivial potential 
even in the supersymmetric case due to the finite temperature 
effect, which breaks the supersymmetry explicitly. 
In the high-temperature limit, from Eqs.~(\ref{26}),
the effective potentials have the following form
\begin{mathletters}
\bea
\label{b26}
\tilde E^{bKK}_4(\beta)&=&
{V_3 \over 2\beta^4}{\pi^2 \over 15} A_0(a)
-{V_3 \over 2 \beta^2}
{ a^2 k^2 \over 12} A_{-2}(a) + {\cal O}\left(1\right)\ , \\ 
\label{b27}
\tilde E^{fKK}_4(\beta)&=&
{V_3 \over 2 \beta^4}{7\pi^2 \over 120} A_0(a)
 - {V_3 \over 2 \beta^2}
{ a^2 k^2 \over 24} A_{-2}(a) + {\cal O}\left(1\right) \ ,\\ 
\label{b28}
E^{sKK}_4(\beta)&=&
{V_3 \over 2 \beta^4} {\pi^2 \over 8} A_0(a)
- {V_3 \over 2 \beta^2}
{a^2 k^2 \over 8} A_{-2}(a) + {\cal O}\left(1\right) \ .
\eea
\end{mathletters}
Using the explicit forms of $A_0(a)$ given in Eq.~(\ref{bb4}) and 
$A_{-2}(a)$ given in Eq.~(\ref{bb9}), the above energies have the 
following form:
\be
\label{b29}
\tilde E^{i KK}_4 (\beta ;a) = {B_0^i \over \beta^4} 
+ {B_1^ik^2 \over \beta^2}\left(1+a^2\right) 
+ {\cal O}\left(\beta^0\right)\ .
\ee
Here $i=b$, $f$, $s$ and $\tilde E^{sKK}_4 =E^{sKK}_4$. 

We now add the contributions from the zero-point energies, 
which were subtracted in Eq.~(\ref{b21}). The contributions were 
evaluated in \cite{GR} and have the 
following form:
\bea
\label{gr1}
&& E^{bKK}_4(\beta=\infty)
= - E^{fKK}_4(\beta=\infty)
= {k^4 a^4 V_3 \over 16\pi^2}
\int_0^\infty dt\,t^3\ln \left[1 - {k_\nu (t) i_\nu (at) 
\over k_\nu (at) i_\nu (t)}\right] \ ,\nn
&& \quad E^{sKK}_4(\beta=\infty)= 0\ .
\eea
Here we neglect terms that can be absorbed into the 
redefinition of the brane tensions.
For small $a$, we have 
\bea
\label{gr2}
\int_0^\infty dt\,t^3\ln \left[1 - {k_\nu (t) i_\nu (at) 
\over k_\nu (at) i_\nu (t)}\right] 
 ={2 \over \nu \Gamma(\nu)^2}\left({\nu - \alpha + 2 
\over \alpha + \nu -2}\right)\left({a \over 2}\right)^{2\nu}
\int_0^\infty dt\,t^{2\nu + 3}{k_\nu (t) \over i_\nu (t)} 
+ \cdots 
\eea
when $\alpha + \nu \neq 2$ and 
\be
\label{gr3}
\int_0^\infty dt\,t^3\ln \left[1 - {k_\nu (t) i_\nu (at) 
\over k_\nu (at) i_\nu (t)}\right] 
={2(\nu -1) \over \Gamma(\nu)^2}
\left({a \over 2}\right)^{2\nu-2}
\int_0^\infty dt\,t^{2\nu + 1}{k_\nu (t) \over i_\nu (t)} 
+ \cdots 
\ee
when $\alpha + \nu = 2$. 
The $a$ dependent part of the total effective action 
has the following form:
\be
\label{gr4}
V^i(a)={B_1^ik^2 \over \beta^2}a^2 + B_2^i k^4 a^{2\mu}\ .
\ee
Here $\mu=\nu + 2$ when 
$\alpha + \nu \neq 2$ or $\mu=\nu + 1$ when 
$\alpha + \nu = 2$. 
If $\mu\neq 1$ and $-{B_1^i \over \mu B_2^i}>0$, 
$V^i$ has non-trivial extremum at
\be
\label{gr5}
a_m^2 = 
\left(-{B_1^i \over \mu k^2\beta^2 B_2^i}\right)^{1 \over \mu -1}\ ,
\ee
which determines the distance between the branes. As an example 
 we consider the case in which $\alpha=2$ and 
$\nu>0$. Then from Eq.~(\ref{b25b}), we find
\be
\label{b34}
i_\nu(z)=zI_\nu'(z)>0\ ,\quad k_\nu(z)=zK_\nu'(z)<0\ ,
\ee
and therefore from Eq.~(\ref{gr1})
\be
\label{b35}
B_2^b=-B_2^f 
={V_3\over16\pi^2} {2^{1-2\nu} \over \nu \Gamma(\nu)^2}
\int_0^\infty dt\,t^{2\nu + 3}{K'_\nu (t) \over I'_\nu (t)} 
<0\ .
\ee
On the other hand, from Eq.~(\ref{bb9}) with $\alpha=2$, 
Eqs.~(\ref{b26})--(\ref{b28}) and Eq.~(\ref{b29}), we find $B_1^i$ is 
negative if $\nu^2>3/4$.
Then, in case of fermions, $V^f$ can have an extremum, 
which is a minimum. In the special case of $\nu={5 \over 2}$, 
as an example,  using numerical integration one gets
\be
\label{b36}
\int_0^\infty dt\,t^8{k_{{5 \over 2}} (t) \over 
i_{{5 \over 2}} (t)} = -1492.97\dots\ .
\ee
Then 
\bea
\label{b37}
{B_1^f \over V_3} &=& -{11\over384}=-0.0286458 \nn 
{B_2^f \over V_3} &=& 0.133752\dots\ ,
\eea
and 
\be
\label{b38}
a_m
= \left({0.0475938 \over k^2 \beta^2}\right)^{1 \over 7}\ .
\ee
Thus, we explicitly demonstrate that quantum spontaneous 
compactification of the 5-dimensional
AdS brane-world at nonzero temperature is 
possible.

Now one can try to estimate the value of temperature which is 
necessary to solve the hierarchy problem.
Since $k^2=(10^{19} \mbox{GeV})^2$, in order that $a_m$ be
the ratio of the weak scale to the Planck scale, $10^{-17}$, 
we find ${1 \over \beta}\sim 10^{-40} \mbox{GeV}\,\sim10^{-27}\,\mbox{K}$. 
This is an extremely low 
temperature. 
Of course, this also means the high temperature 
expansion is not valid since the dimensionless expansion 
parameter is $k\beta\sim 10^{59}$.

\section{Low-Temperature Limit}

Thus, we are led to consider the low temperature limit. We start from 
Eq.~(\ref{b21c}). By changing the variable from $q$ to $s$ 
\be
\label{l1}
q=\sqrt{s^2 + 2\beta ms}\ ,
\ee
the energies $\tilde E^b_d(\beta)$ and $\tilde E^f_d(\beta)$ 
for $d=4$ in Eq.~(\ref{b21c}) have the following forms:
\bea
\label{l2}
\tilde E^{b,f}_4(\beta)={V_3 \over 2\pi^2\beta^4}
\int_0^\infty ds {\left( s+\beta m \right)^2 
\sqrt{s^2 + 2\beta ms} \over \e^{s+\beta m} \mp 1} \ .
\eea
Then in the low temperature limit, where $\beta\rightarrow 
\infty$, one gets
\be
\label{l3}
\tilde E^{b,f}_4(\beta)\rightarrow
{V_3 m^{5 \over 2}\e^{-\beta m} \over \sqrt{2}\pi^2\beta^{3 \over 2}}
\int_0^\infty ds s^{1 \over 2}\e^{-s} 
={V_3 m^{5 \over 2}\e^{-\beta m} \over (2\pi \beta)^{3/2}}
\ee
We now consider the sum of the Kaluza-Klein modes, but due to 
the factor of $\e^{-\beta m}$, when $\beta$ is large
we need to include only the 
lowest root of Eq.~(\ref{b9}), $x=x_1$.  In general $x_1$ 
depends on $a$ but Eq.~(\ref{b9}) reduces to
\be
\label{l4}
-(2-\alpha - \nu )\Gamma(\nu)
\left({ax_1 \over 2}\right)^{-\nu}
j_\nu(x_1)=0
\ee
when $a$ is small. Then $x_1$ satisfies 
$j_\nu(x_1)=0$ ($\alpha+\nu\ne2$) in the limit and one can regard 
that $x_1$ is of order unity. Adding the contribution 
from the zero-point energies as in Eq.~(\ref{gr4}), we find the 
following effective potential 
\bea
\label{l5}
V^i(a)&=& B_2^i k^4 a^{2\mu} + B_3^i\beta^{-{3 \over 2}}
(ka)^{5 \over 2}\e^{-\beta ka x_1} \nn
&=&k^4B_2^i\left(a^{2\mu}
+ {B_3^i\over B_2^i} (\beta k)^{-3/2}a^{5/2}\e^{-\beta ka x_1}\right) \ .
\eea
Here 
\be
\label{l6}
B_3^b=B_3^f = {B_3^s \over 2}=
{V_3 x_1^{5 \over 2} \over (2\pi)^{3/2} }>0\ .
\ee
Since $B_3^i$ is positive, the second term in the effective 
potential (\ref{l5}) has a maximum when $\beta k a\sim 1$ and 
exponentially approaches zero after the maximum. Then if 
$\beta k$ is large, the effective potential (\ref{l5}) 
has a nontrivial minimum if $\mu >5/4$ and $B_2^i$ is 
positive. Thus when $\beta k$ is large, 
the order of magnitude of the minimum $a_m$ in the effective potential 
(\ref{l5}) is given roughly by
\be
\label{l7}
a_m\sim {1 \over \beta k}\ln \beta k \ .
\ee
Then since $k\sim 10^{19}\mbox{GeV}$, 
if ${1 \over \beta}\sim 10\,\mbox{GeV}$, we have 
$a_m\sim 10^{-17}$ and the weak scale can be generated : 
$a k \sim 10^2 \,\mbox{GeV}$. 

Of course, the global minimum of the potential (\ref{l5}) occurs at
$a=0$, while the minimum at $a=(\beta k)^{-1}\ln\beta k$ is only local.
But the barrier height is very high.  We estimate the tunneling
probability as proportional to the WKB factor
\begin{equation}
\exp\left[-2\int dx\,\sqrt{2m(V-E)}\right]\sim\exp\left[-\sqrt{mV_{\rm max}}
L\right].
\end{equation}
The numerical value here is composed of
\begin{equation}
\sqrt{V_{\rm max}}\sim(1/\beta^2)\sqrt{V_3}\sim 10^2\,\mbox{GeV}^2(10^{28}\,
\mbox{cm})^{3/2}\sim 10^{65}\,\mbox{GeV}^{1/2}
\end{equation}
and
\begin{equation}
\sqrt{m}=\sqrt{kax}\sim10\,\mbox{GeV}^{1/2}.
\end{equation}
So even with $L$ as small as the Planck scale, $L\sim 10^{-19}\,
\mbox{GeV}^{-1}$, we have a negligible tunneling probability.

As an example of how this minimum can be achieved,
we consider the case that $\nu>0$ and $\alpha=2$. 
Then because $\mu=\nu+2>2$ and $B_2^f>0$ from Eq.~(\ref{b35}), 
we have the minimum of the potential for the fermionic case. 
If $\nu=5/2$ or $\mu=9/2$, $x_1=3.6328$, $B_3^f/B_2^f=11.9408$, and
we find the minimum occurs at $a_m=5.15\times 10^{-17}$.
In other words, a bulk quantum fermion may generate 
a thermal (flat) 5-dimensional AdS brane-world with 
the necessary hierarchy scale! 
This example proves that quantum bulk 
effects in a brane-world  AdS$_5$ at nonzero temperature may 
not only stabilize the brane-world (quantum spontaneous 
compactification occurs) but also provide
the dynamical mechanism for the resolution of the hierarchy problem 
(with no fine-tuning). Note, however, that the temperature in the above 
example is less than the characteristic temperature in a hot 
inflationary universe (but is of the order of the Hubble temperature).
 It is very important to recognize that 
despite the fact that our toy example involved only bulk matter 
fields it is expected that such a dynamical mechanism 
occurs in terms of the AdS/CFT correspondence, due to the 
contribution of thermal 5-dimensional gravitons and matter 
supermultiplets which appear after sphere compactification 
of IIB supergravity on the AdS$_5$ space.

The example for the single fermion is somewhat unsatisfactory because
the minimum at $a\ne0$ is only local.  A more elaborate model consists
of $N$ bosons and $M$ fermions.  For simplicity we take all the bosons
to have a common mass, and hence a common power $\mu_b$, and similarly
that the fermions have a common power $\mu_f$.  First we consider the
case of zero temperature.  Then we have the following effective potential,
for $a\ll1$,
\begin{equation}
V_{\beta=\infty}(a)=NB_2^bk^4a^{2\mu_b}+MB_2^fk^4a^{2\mu_f}.
\label{bfep0}
\end{equation}
When $\alpha=2$, we have $B_2^b<0$ and $B_2^f>0$.  We will also assume that
$\mu_b<\mu_f$.  Then the potential is negative when $a$ is small because
the contribution from the boson [the first term in Eq.~(\ref{bfep0})]
dominates.  On the other hand, the potential becomes positive when $a$ is
large because the contribution from the fermion [the second term in 
Eq.~(\ref{bfep0})] dominates.  Therefore, the potential has a global
minimum for $a\ne0$ if $\mu_b<\mu_f$.  Since
\begin{equation}
{dV_{\beta=\infty}\over da}=2\mu_bNB_2^bk^4a^{2\mu_b-1}
+2\mu_fMB_2^fk^4a^{2\mu_f-1},
\end{equation}
the minimum occurs at
\begin{equation}
a=a^0_{\rm min}=\left(-{\mu_bNb_2^f\over\mu_fMB_2^f}\right)^{1\over
2(\mu_f-\mu_b)}\sim{\cal O}(1).
\end{equation}
We should note that $0<a<1$ because the distance between the branes is
$r_c=-(1/k\pi)\ln a$.  Thus, only if $a^0_{\rm min}<1$ do we have a 
meaningful solution. (Of course, our equations are only valid if $a\ll1$.)
 Now we include the effect of finite temperature,
in particular the low-temperature effective potential ($a\ll1$)
\begin{equation}
V(a)=NB_2^bk^4a^{2\mu_b}+MB_2^fk^4a^{2\mu_f}+(N+M)B_3\beta^{-3/2}(ka)^{5/2}
e^{-\beta kax_1}.
\label{bfeffpot}
\end{equation}
Here $B_3=B_3^b=B_3^f$.  First we consider the case when $\beta$ is not
many orders of magnitude larger than $1/k\sim 10^{-19}\,\mbox{GeV}^{-1}$,
but still satisfying $k\beta\gg1$ so that the low temperature approximation
remains valid.  The last term in Eq.~(\ref{bfeffpot}) shifts the potential
and makes two local minima, say $a_1<a_2<1$.  If $\mu_b<5/4$, we have $a_1
\ne0$.  As the temperature decreases $a_1$ approaches zero.  For high
temperature, the corresponding minimum is a global one, but this ceases to
be true when $\beta^{-1}\sim 1\,\mbox{GeV}$.  When $\mu_b>5/4$,
evidently $a_1=0$, that is, the distance between the two branes is
infinite.  Again, for high temperature, the minimum at $a=a_1=0$ is a
global one, but at low temperature the global minimum lies at $a=a_2\approx
a^0_{\rm min}$.  This is the sketch of an argument that at some critical
temperature the global minimum shifts to a small, but nonzero, $a$.  This
then suggests that the thermal RS universe may be created as the result of
a phase transition; the phase where symmetry breaking
of the potential occurs is stable.
\section{Discussion}

In summary, we have formulated quantum bulk scalar (or spinor) dynamics in 
a brane-world 
AdS$_5$ at nonzero temperature. Such a universe is a natural 
generalization of the Randall-Sundrum scenario.
The calculation of the thermodynamic energy (effective potential) is performed
in the high- and low-temperature limits. Thermal quantum corrections generate 
a modulus potential which in some explicit examples has been shown to have a
minimum.
Hence, a new dynamical mechanism to stabilize the thermal brane-world 
universe has been suggested. Moreover, the finite separation of flat branes 
may be fixed in terms of the Planck constant and the
 temperature so that the hierarchy problem is naturally solved.

 The unsatisfactory feature of our mechanism is that although quantum 
 spontaneous 
compactification occurs, the natural hierarchy is generated at quite small 
values of temperature, much smaller than the temperature 
which is characteristic for a hot inflationary universe. The good feature 
of our proposal is that it is directly applicable to the RS warped 
compactification in terms of the AdS/CFT correspondence. The reason is that 
the required thermal modulus potential may be easily generated by bulk quantum 
supergravity. Only the numerical coefficients in the potential (\ref{l5}) (or 
Eq.~(\ref{gr4})) will be changed due to the contribution of the
thermal graviton (and superpartners). 
It remains to be checked that these explicit values correspond to the
extremum of graviton potential being a minimum, a maximum, or an inflection
point.

In any case the possibility of obtaining a thermal modulus potential 
with a possible extremum opens a new perspective on the
dynamical formulation of
brane-worlds. So far the only known (classical) dynamical mechanism to 
solve the hierarchy problem has been proposed in Ref.~\cite{mod}. Unlike
our scheme, that mechanism
does not permit the extension to bulk gravity (AdS/CFT correspondence).   

As a final remark let us note that it would be very interesting 
to generalize our scenario to the case where the branes are curved (but with no 
changes in the bulk space). It could  perhaps result
in the construction of a new inflationary brane-world at nonzero 
temperature where thermal quantum effects generate not only inflation
 (as in case of the nonthermal scenario of Refs.~\cite{NOZ,HHR}) but also
a hierarchy of scales corresponding to a physically reasonable hot 
universe.\footnote{For another study of the thermodynamics of brane
worlds, with empahsis on thermalization of black holes in the bulk
with the brane, see Ref.~\cite{CKN}.}

\section*{Acknowledgements}
The work of KAM was supported in part by a grant from the 
US Department of Energy. The work of SDO was supported in 
part by CONACYT (CP, Ref.~990356 and grant 28454E) and in 
part by RFBR grant N99-02-16617.

\end{document}